# DID CHRISTIAAN HUYGENS NEED GLASSES? A STUDY OF HUYGENS' TELESCOPE EQUATIONS AND TABLES

by

ALEXANDER G. M. PIETROW[1,2,*]

[1]*Institute for Solar Physics, Department of Astronomy, Stockholm University, Albanova University Centre, SE-106 91 Stockholm, Sweden*
[2]*Leibniz-Institut für Astrophysik Potsdam (AIP), An der Sternwarte 16, 14482 Potsdam, Germany*

In the later stages of his life, Christiaan Huygens semi-empirically derived a set of relations between the objective focus and diameter, the eyepiece focus, and the magnification that resulted from combining the two lenses. These relations were used by him and his brother to build what he believed were optimized telescopes. When comparing these equations to the ones derived from modern optical principles, Huygens' telescopes were in fact far from optimal. While there are several potential reasons for this discrepancy, one possible reason, explored in this work, is that Huygens might have suffered from a mild case of myopia (or near-sightedness) and that he compensated this condition by building telescopes that overmagnified by a factor of 3.5. Based on this hypothesis, Huygens' visual acuity is estimated to be around 20/70, which on average corresponds to an optical prescription of −1.5 diopters.

## INTRODUCTION

Christiaan Huygens was a seventeenth-century polymath known for revolutionizing the fields of optics, mechanics, time keeping and astronomy. In the latter, he made several discoveries, including the true nature of Saturn's rings and its moon Titan. Huygens designed and produced his own lenses, often with the help of his brother Constantijn. At that time both optical theory and calculus were still at an early stage of development, meaning that the theory of refraction and the material physics of glass were still largely misunderstood.[1] This, coupled with the different types and quality of glass that were available throughout Europe, made it difficult to disentangle theoretical effects from material ones, resulting in

*apietrow@aip.de

1  John Dollond and James Short, 'An account of some experiments concerning the different refrangibility of light. By mr. John Dollond. With a Letter from James Short, M. A. F.R.S. Acad. Reg. Suec. Soc.', *Phil. Trans.* **50**, 733–743 (1757).

many of the found relations being largely based on trial and error.[2] For example, Adrien Auzout published a table in 1665 where he gives three values for the object glass diameter[3] $D$ versus the focus $f$ based on the quality of said lens.[4] The lens production process of the Huygens brothers worked in much the same way, and was perfected by trial and error as well as the study of other lenses, optical theory and the methods of other lens makers.[5]

When Huygens started to make long focus lenses, he was aware of a quadratic relation between the size of a lens and its focus, as dictated by the discovery of chromatic aberrations by Newton.[6] In either 1684 or 1685[7] Huygens wrote down his telescope equations in the form of a semi-empirical table that related the objective focus, the objective aperture, the eyepiece focus and the magnification of Keplerian refractors[8] at various focal lengths. In this table, he started with a telescope objective and eyepiece pair for which he knew the magnification and then calculated matching values for other objective foci, as can be seen in the description of the original table:[9]

> … *Quibus astronomico et mathematico calculo probatis. Hanc tabulam subjunxit. In qua ex amplificatione, quae est in 30 pedum telescopio ut 109 ad 1, caeterae omnes ex proportione aperturarum repertae sunt, et illam relatae* …
>
> [English: … Having proved these by astronomical and mathematical calculation. He attached this table, in which he gives the magnification, which for a 30-foot telescope is found to be as 109 to 1. All the rest in proportion to the openings they were derived, and related to this value … ]

This 30-foot telescope is also mentioned several times in volume 13 of his *Oeuvres Complètes* (OC); for example, on page 494 where Huygens states that [English transl.] 'experience teaches us that a lens of 30 feet fits an aperture of 3 inches'. Furthermore, footnotes 2 and 5 on page 497 show that in his draft Huygens experimented with different objective and eyepiece focus values, both of which resulted in a higher magnification for his 30-foot telescope, before changing his mind and using the values given above. These values were optimized by Huygens for the observation of Saturn, as stated in the first footnote of page 498, where it is added that smaller openings can be used for brighter objects and that for daytime use the focus should be doubled (thus halving the magnification).

By studying the optical design equations that Huygens adhered to while designing his telescopes, and by comparing them to their modern counterparts, it is possible to shed light on the limitations of seventeenth-century telescope design. Additionally, by assuming that the aforementioned 30-foot objective and eyepiece pair were manually optimized by Huygens until he achieved the best magnification and clarity, it becomes possible to reverse-engineer the situation and learn about his eyesight by comparing the capacity of the telescope to the ideal case.

---

2 Audouin Dollfus, 'Christiaan Huygens as telescope maker and planetary observer', in *Proceedings of the international conference 'Titan: from discovery to encounter'*, 13–17 April 2004, ESTEC, Noordwijk, Netherlands (ed. Karen Fletcher; ESA SP-1278), pp. 115–132 (2004).

3 Throughout this work we mean the clear diameter after the aperture stop is applied, not the total diameter of the lens. See figure 15 of Audouin Dollfus, 'Les frères Huygens et les grandes lunettes sans tuyau', *L'Astronomie* **112** (1998) at p. 128, and Frederik Kaiser, 'Iets over de kijkers van Christiaan en Constantijn Huygens', *Instituut* **VI**, 417 (1846).

4 Adrien Auzout, 'Monsieur Auzout's judgment touching the apertures of object-glasses, and their proportions, in respect of the several lengths of telescopes', *Phil. Trans.* **1**, 55–56 (1665) (http://www.jstor.org/stable/101425).

5 Peter Louwmam, 'Christiaan Huygens and his telescopes', *op. cit.* (note 2), pp. 103–114 (2004).

6 Isaac Newton, 'A letter of Mr. Isaac Newton, Professor of the Mathematicks in the University of Cambridge; containing his new theory about light and colors: sent by the author to the publisher from Cambridge, Febr. 6. 1671/72; in order to be communicated to the R. Society', *Phil. Trans.* **80**, 3075–3087 (1671). Christiaan Huygens, *Astroscopia compendiaria: tubi optici molimine liberata* (Leers, The Hague, 1684).

7 Christiaan Huygens, *Oeuvres Complètes de Christiaan Huygens*, vol. 13 (M. Nijhoff, The Hague, 1899) at p. 497.

8 Telescopes that consist of a simple positive objective lens and a simple positive eyepiece lens. They produce a relatively large field of view but invert the image, making them more popular for astronomical observations than terrestrial viewing.

9 Codices Hugeniani (HUG) volume 32, fols 030r–v. (n.d.). Brill: https://doi.org/10.1163/2468-0303-cohu_32-011.

In this article, the telescope equations that can be extracted from the table and those found in his writings are compared to the ones derived from modern optical principles, and potential causes for discrepancies between the two are investigated with an emphasis on Huygens' eyesight.

## EQUATIONS FROM TEXT

As calculus was still in its early stages, the equations by Huygens are written down verbally, often in a rather roundabout way. In this section these equations are given in their original language, a rough translation to English, and transcribed into a modern notation. The first of these equations can be found in a letter by Huygens to his brother Constantijn, sent on 23 April 1685,[10] where Huygens states:

> *Et pour trouver l'un ou l'autre j'ay cette regle aisée, qui est de multiplier les pieds de la longueur du verre objectif par 3000, et de tirer la racine quarrée du produit, laquelle marque les pouces dixiemes et centiemes de l'ouverture et en mesme tems du foier de l'oculaire.*
>
> [English: Multiply the feet of the length of the objective glass by 3000, and then draw the square root of the product, which marks the tenths and hundredths inches of the aperture and at the same time the eyepiece length.]

This can be written as an equation in the form:

$$\sqrt{3000 \cdot f_o[\text{Rhineland feet}]} = 100 D [\text{Rhineland inch}] \text{ and } f_e = D$$

This equation can then be simplified and the units[11] converted to centimetres to obtain the well-known equation that was reported by Dollfus in 1998[12] as well as a relation between lens diameter and eyepiece focus:

$$f_o = 15.3[\text{cm}^{-1}] D^2$$

$$f_e = 1[\text{cm}^{-1}] D$$

However, in his *Dioptrica* the text is slightly different, with the last sentence instead reading:[13]

> *Eadem si augeatur decima sui parte, dabit foci distantiam lentis ocularis iisdem centesimis expressam.*
>
> [English: The same, if it be increased by a tenth of itself, will give the focal distance of the lens of the eye, expressed in the same units.]

Thus, from this text the equation has the following shape:

$$f_e = 1.1[\text{cm}^{-1}] D.$$

---

10   Huygens, *op. cit.* (note 7), vol. 19, at p. 7.
11   An old Dutch unit of length where 1 Rhineland foot is equal to 31.4 cm. See Alex Pietrow, 'De Rijnlandse Roede, de geschiedenis van een oud-Hollandse lengtemaat', *Nederlands Tijdschrift voor Natuurkunde*, **83** (3), 74–76 (2017).
12   Dollfus, *op. cit.* (note 3), at p. 126.
13   Christiaan Huygens, *Opuscula posthuma* (Cornelius Boutesteyn, Netherlands, 1703), at p. 210.

This difference can also be found in the tables, where the original manuscript and the *Dioptrica* use the latter values, while the aforementioned letter and the OC uses the former. The lower value for the eyepiece focus corresponds to a larger magnification.

Returning to the letter, we find that Huygens continues with the following:

> *voicy l'autre regle pour scavoir combien de fois la lunette multiplie selon le diametre. Il faut multiplier les pieds de la longueur par 480, et la racine quarrée du produit sera le nombre de la multiplication ou grossissement qu'on cherche.*
>
> [English: To know how many times the objective multiplies according to the diameter, we must multiply the focal length in feet by 480. The square root of the product will be the magnification that we seek.]

This can be rewritten as:

$$\sqrt{480 \cdot f_o[\text{Rhineland feet}]} = M,$$

which once again can be simplified to:

$$f_o = 0.07[\text{cm}^{-1}]M^2.$$

We can find one more equation on page 162 of the *Dioptrica*:

> *Five quae nudo oculo percipitur est ea quae foci distantiae lentis e terioris ad foci distantiam ocularis.*
>
> [English: The distance perceived by the naked eye is that of the focal distance of the lens from the posterior to the focal length of the eyepiece.]

This is the same relation that is used today:

$$M = f_o/f_e.$$

It is also worth mentioning that Huygens mentions in the same letter that these equations are to be used as a rule of thumb for when the table is not at hand, possibly implying that these equations are less accurate than his tables. In the next section the focus will shift to tables, and a comparison will be made between these equations and the ones extracted from the table.

## Equations from the tables

As mentioned in the prior section, two versions of the telescope tables were prepared by Huygens. The older one can be found in the *Codices Hugeniani*[14] and on page 211 of the *Dioptrica*, and a more recent one can be found in the OC. The main difference between the two is whether the focus is equal to the aperture diameter or whether it is scaled with the aforementioned value of 1.1, although the other numbers do change because of this. One potential reason for this change is that it makes it simpler to work with the various parameters, as well as that it gives a better performance for the telescopes (e.g. larger magnification).

The equations in table 1 here can be found by applying a linear or a quadratic fitting routine (where appropriate) to these tables.[15] Looking at the two columns, it immediately becomes clear that the latter is different because $f_e = D_o$. The equations extracted from the tables are very similar to the verbal equations, which shows the precision with which Huygens worked. However, it is important to note that the equations that deal with the magnification in the left column can significantly change depending on the precision of the constants. This problem does not manifest in the right column as most of the constants become equal to one another. This could have been the reason for equating the eyepiece focus with the objective diameter.

---

14   HUG 32, *op. cit.* (note 9).
15   The tables and code used to create these fits are available on the project Github (https://github.com/AlexPietrow/Huygens).

Table 1. Huygens' telescope equations as he first calculated them as well as an updated set. All constants are given in units of [cm$^{-1}$].

| *Codices Hugeniani* | *Oeuvres Complètes* |
|---|---|
| $f_o = 15.29 D_o^2$ | $f_o = 15.29 D_o^2$ |
| $f_o = 12.84 f_e^2$ | $f_o = 15.29 f_e^2$ |
| $f_o = 0.08 M^2$ | $f_o = 0.07 M^2$ |
| $f_e = 1.09 D_o$ | $f_e = 1.00 D_o$ |
| $M = 13.89 D_o$ | $M = 15.29 D_o$ |
| $M = 12.74 f$ | $M = 15.29 f$ |

The question now is how well Huygens understood the limitations of his optics, and if he truly reached the optimal performance. An attempt can be made to answer these questions after the modern counterparts of these equations are derived.

## EQUATIONS FROM MODERN OPTICS

Our understanding of optics has vastly improved since the time of Huygens. The development of calculus, along with an understanding of geometrical and diffractional optics, allows us to derive some of these equations from fundamental principles rather than doing so empirically. In this section we will do so where possible and compare Huygens' telescope equations to the true limit of a singlet lens to see how well his telescopes perform under a modern standard.

We begin with the first equation that relates the objective focus to its diameter. We can derive this relation by equating the depth of focus[16] to the chromatic aberration[17] so that all colours are within this distance. From a textbook[18] we find the former to be:

$$L_f = 2.4 \lambda f_o^2 / D^2$$

with the depth of focus $L_f$, the wavelength $\lambda$ and $f_o$ and $D$ in metres.

The chromatic aberration can likewise be found in a textbook,[19] where we find that for lenses with a long focal length the aberration is given by:

$$L_{ch} = f_o / V$$

with $V$ the Abbe number.[20]

---

16  The lateral tolerance to the placement of the focal plane of a lens. Typically, the Rayleigh criterion is used for this, which states that the image quality is not sensibly degraded as long as the defocus is below $\lambda/4$.

17  Singlet lenses cannot focus all colours on the same point, which means that one can only focus on one colour and has to deal with a defocused image in the other colours.

18  Ariel Lipson, Steven G. Lipson and Henry Lipson, *Optical physics* (Cambridge University Press, 2010) at p. 60.

19  Rudolf Kingslake and Barry Johnson, *Lens design fundamentals* (Academic Press, London, 2010) at p. 162.

20  The Abbe number is an approximate measure of the dispersion characteristics of the glass. It describes how much the refractive index changes with wavelength, with high values of V indicating low dispersion.

We then equate the two derived distances and obtain our relation between aperture and focal length in the form:

$$f_o = D_o^2/2.4\lambda V.$$

We can see that this equation is of the same form as the first telescope equation of Huygens.

However, if we fill in the numbers[21] ($\lambda = 550$ nm and $V = 60$, as found by Mills and Jones[22]) to find the constant, we get a factor of roughly 126.3 cm$^{-1}$, instead of Huygens' 15.2 cm$^{-1}$. To make the two factors match we have to change our initial assumption for the lateral tolerance being $\lambda/4$, as mentioned in footnote 16. By allowing a larger difference in path length, and thus defocus, we can get a smaller constant. In this case, we need a roughly five times larger path length ($2.5\lambda$).

This suggests that Huygens' lenses were purposefully designed sub-optimally with a shorter focus than is necessary to fully remove the aberrations based on the aperture diameter. It is possible that this was done out of necessity to keep telescope lengths manageable, as aerial telescopes were already incredibly long.[23] Without compromising, a telescope with a focus of 505 m would be needed to correct for a telescope with a diameter of 20 cm (versus the roughly 60 m lengths that Huygens required).

The spherical aberration[24] is not included in these calculations. This is because this aberration does not significantly degrade the resolving power of a telescope with a high $F$-number,[25] even though it was believed to be the main cause of defocus in Huygens' time. In fact, it was only after Newton discovered the spectral nature of light that it became apparent that spherical and chromatic aberration were found to be different phenomena. The negligible nature of spherical aberration in high $F$-number telescopes can be illustrated by the much more modern Swedish Vacuum Solar Telescope[26] ($D = 0.5$ m, $f = 22.35$ m, F/50) that had a negligible amount of spherical aberration, versus the Swedish 1-m Solar Telescope[27] ($D = 1$ m, $f = 20.3$ m, F/20) that has a much larger amount that needs to be dealt with using adaptive optics.

Next we derive the fifth telescope equation that relates the aperture diameter to the maximal useful magnification, as this will aid us in obtaining the other relations. We start with the equation that describes the maximal resolution of a lens:

$$\alpha = 1.22\lambda/D$$

with $\alpha$ in radians, and $\lambda$ and $D$ in metres.

---

Table 2. Huygens' telescope equations and the same equations calculated from modern optical principles. For convenience, we also give the *F*-number and exit pupil (EP)[28] size for each column. All constants except the EP are given in units of [cm$^{-1}$].

| Codices Hugeniani | *Oeuvres Complètes* | Modern optics |
|---|---|---|
| $f_o = 15.29 D_o^2$ | $f_o = 15.29 D_o^2$ | $f_o = 126.3 D_o^2$ |
| $f_o = 12.84 f_e^2$ | $f_o = 15.29 f_e^2$ | $f_o = 0.15 f_e^2$ |
| $f_o = 0.08 M^2$ | $f_o = 0.07 M^2$ | $f_o = 2.42 M^2$ |
| $f_e = 1.09 D_o$ | $f_e = 1.00 D_o$ | $f_e = 29.2 D_o$ |
| $M = 13.89 D_o$ | $M = 15.29 D_o$ | $M = 4.33 D_o$ |
| $M = 12.74 f_e$ | $M = 15.29 f_e$ | $M = 0.15 f_e$ |
| F# = 15.29 $D_o$ | F# = 15.29 $D_o$ | F# = 126.3 $D_o$ |
| EP = 0.7 [mm] | EP = 0.65 [mm] | EP = 2.3 [mm] |

This equation can be simplified by filling in 550 nm for the wavelength, giving us

$$\alpha_m = 0.23 [\text{arcmin cm}]/D$$

with $\alpha_m$ in arcminutes and $D$ in centimetres.

A person with 20/20 vision has a visual acuity of 1 arcminute,[29] meaning that this is the smallest angle separating details that can be discerned. We can use this to get the maximal useful magnification[30] by requiring that the smallest resolution element times the magnification is equal to this number. We can then substitute this value into $\alpha_m$ and get

$$M = 4.33 [\text{cm}^{-1}] D_0$$

with $D$ in cm.

We can see that this is more than a factor three smaller than the values in either the new or the old telescope equations, demonstrating that Huygens substantially overmagnified.

We can then obtain the remaining equations for the minimal eyepiece length from the relation above after we insert $M = f_o/f_e$, which gives us $f_e = f_o/4.33 D_o$. We can then

---

28  Exit pupil is the size of the beam leaving the telescope eyepiece. It is around 0.7 mm for Huygens' telescopes and just over 2 mm for the ideal case. This means that the eye must be very precisely placed in order to see anything at all through Huygens' telescopes. For comparison, telescopes and binoculars that are easy to look through have exit pupils of about 10 mm, or slightly larger than the maximum size of the observer's pupil, to allow for the head to move while observing. Most will also advise against using an exit pupil below 0.5 mm, as you start seeing the inside of your eye. See, e.g., W. Sheehan and T. Dobbins, 'The spokes of Venus: an illusion explained', *J. Hist. Astron.* **34** (1), 53–63 (2003).

29  H. Snellen, *Probebuchstaben zur Bestimmung der Sehschärfe* (Van de Weijer, Berlin, 1862).

30  This is called the maximal useful magnification, as magnifying beyond this point will not add any additional information and, instead, just blow up the image—on top of that the surface brightness of the whole will go down, making it harder to discern details.

insert the relation for either the objective focus or the objective diameter to obtain

$$f_e = 29.2 D_o$$
$$f_e = 6.74 M$$
$$f_o = 0.15 f_e^2$$
$$f_o = 2.42 M^2$$

with all constants given in units of $[\text{cm}^{-1}]$.

In table 2 we summarize the equations presented in the last two sections. From these equations, the eyepiece focal lengths appear much shorter than necessary. Using longer eyepieces should not have been an issue for Huygens, especially if he used the objective lenses of his small telescopes,[31] which means that it was certainly within his means to improve the performance of his longest telescopes. Huygens' objective/eyepiece combinations led to an overmagnification factor of about 3.5.

These equations are plotted in figure 1, to better illustrate the differences between Huygens' telescope equations and our modern ones.[32] The 1665 'average', 'good' and 'excellent' lens curves derived by Auzout[33] are added to the first panel.

The Huygens curve follows very closely with Auzout's 'good' curve, implying that Huygens' lenses were not exceptional compared to the technology of 20 years prior. This is surprising, given that the Huygens are widely known as being superior for their time. However, these two facts are reconciled if we once again assume that Huygens built his telescopes sub-optimally despite having superior lenses. This seems to be in line with a comparison that Frederik Kaiser made in 1846,[34] where he described the lenses as pristine but stated that Huygens' telescopes had a noticeably inferior resolving power over contemporary achromats, even those with a smaller aperture size, despite the image contrast being better.

Another option is that his lenses were, in fact, dominated by polishing errors and did not reach diffraction limited performance. While proper wave front measurements are needed to confirm this assumption, it is widely accepted that the lenses manufactured by Huygens are of excellent quality and ahead of their time. Additionally, wave front measurements have been performed on several lenses by Campani, and it was shown that most of the objectives were in fact good enough to produce a diffraction limited spot.[35] In this work it is therefore assumed that Huygens' lenses are the same, and that the poorer performance comes purely from the mismatch of objectives and eyepieces.

---

31  A. C. van Helden and R. H. van Gent, 'The lens production by Christiaan and Constantijn Huygens', *Ann. Sci.* **56** (1), IV (1999).
32  Huygens had made a mistake in one of his points in the red curve that he later corrected in the green curve.
33  Auzout, *op. cit.* (note 4).
34  Kaiser, *op. cit.* (note 3).
35  J. Lozi, J. M. Reess, A. Semery, E. Lhomé, S. Jacquinod, M. Combes, P. Bernardi, R. Andretta, M. Motisi, L. Bobis and E. Kaftan, 'Could Jean-Dominique Cassini see the famous division in Saturn's rings?', *SPIE Proc.* **8864**, 88641M (2013).

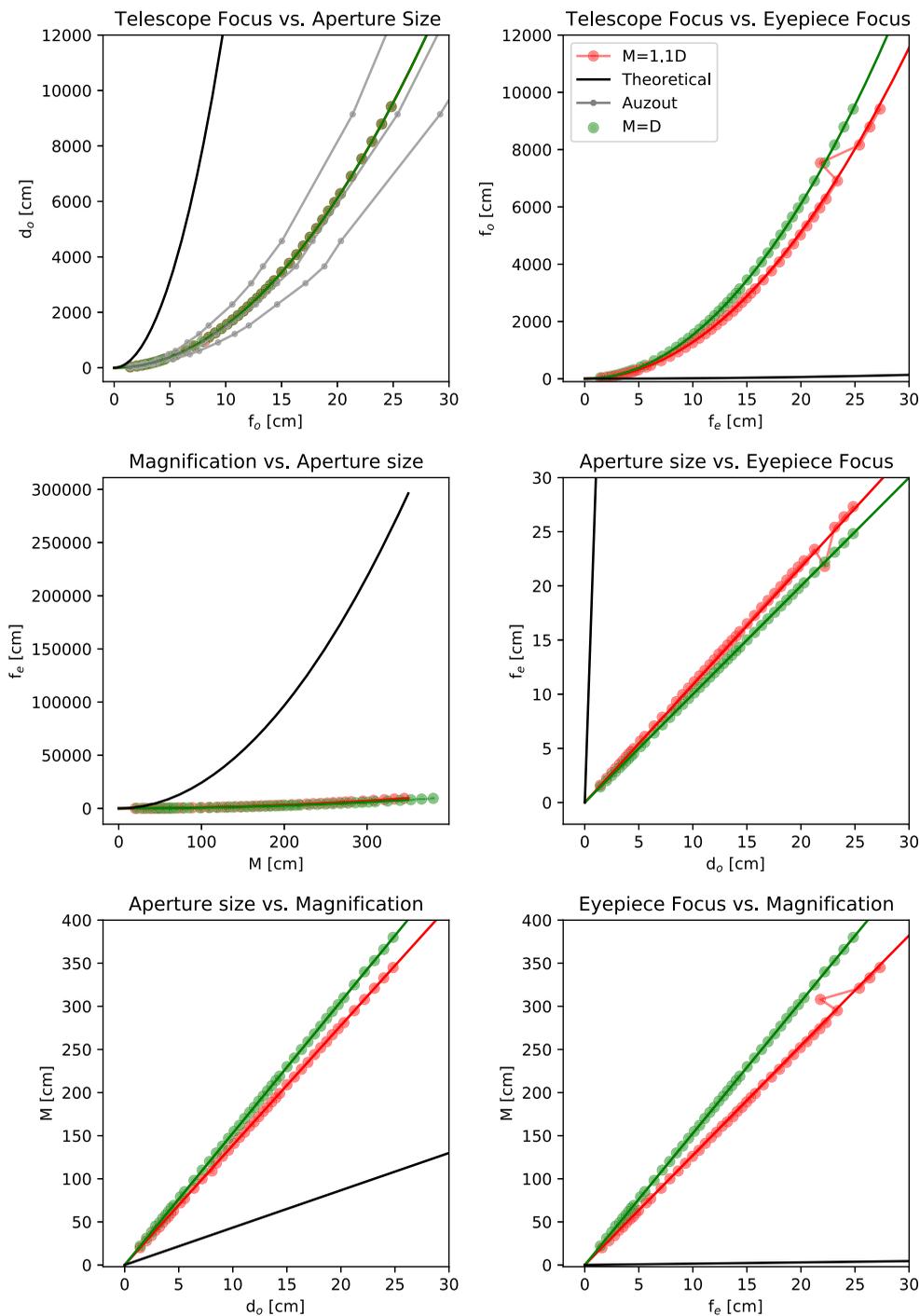

Figure 1. Huygens' telescope equations (red and green) compared to the theoretical limit (black) for the objective focus, objective diameter, eyepiece focus and magnification. Values from a table by Auzout (*op. cit.*, note 4) are included in the first panel, showing that the Huygens curve follows what he would call the trend of 'good' lenses.

## The eyesight of Huygens

When comparing these equations, it is clear that something is strange about Huygens' telescopes and the way that he built them. It is possible that Huygens purposefully chose shorter foci to have manageable lengths for his telescopes or that the seeing in general was so bad[36] that there was no point in going further. However, even though Huygens left a detailed account of his work and motivations, he left no (known) mention of these limiting factors. Yet, he was clearly willing to push the limits of instrumentation to get the clearest image and best magnification, so much so that he was willing to deal with a very small exit pupil, while according to our equations, he could get a better performance and eye relief with longer eyepiece foci. Hence, it is reasonable to assume that Huygens truly believed that his pairing of singlet lenses was the optimal combination.

This seemingly arbitrary limitation could be explained if Huygens had a visual condition and built his telescopes in such a way that he could compensate for it—not unlike people do today when taking off their glasses to look through a telescope and then adjusting the focus to get an image that is sharp for their eyes, a process in which they are in fact replacing the correction of their glasses for an adjustment of the focal length of the telescope.

By assuming that this is the case, it becomes possible to estimate the maximal resolution that Huygens could discern while looking at infinity, which can be easily done by comparing the fifth telescope equation of each column. This is because we know that, in the case of the theoretical example, we assumed a visual acuity of one arcminute to get $M = 4.33 D_o$. However, in order to get the constant to one of the other two values we would need a maximal resolution of 3.51 arcminutes. On a Snellen chart this would correspond to a visual acuity of 20/70, which roughly[37] translates to a refractive error of −1.5 diopters.

This method is an upper limit where we assume that Huygens' eyes are fully responsible for the difference between his equations and the theoretical limits. However, since it is well documented that his father suffered from severe myopia,[38] as well as other members of the family,[39] and considering that myopia has a hereditary tendency, it is reasonable to expect that Huygens might have suffered from this eye condition to some degree. A lower severity of the condition would also explain why he did not consider needing glasses or discussing his condition. He would in all cases be able to read and write and, in a very different world from our modern one, he would also do most other things without great issue.

Another interesting observation is that Huygens has a linear relation between the eyepiece focus and the diameter, one that matches throughout the entire curve. This suggests that Huygens did not in fact empirically measure all the telescopes in his tables, but rather used one, or an average of several to find the basic rules and apply them to all the different objective focal lengths. This is because if he had made multiple measurements, then the constant between the eyepiece focus and the diameter would be different each time unless he had perfect eyesight, given that the relation is no longer linear if myopic effects are considered. We find strong evidence for the former proposition in his notes[40] as well as by the fact that he only mentions the 30-foot telescope in his *Dioptrica*.

---

36 The seeing in the Netherlands, especially Leiden, is exceptionally bad, with around 3 arcseconds on average. A good observatory nowadays aims for 0.5–1.0 arcsecond, e.g. D. Van der Werff, 'Characterization of the atmosphere above', bachelor thesis, Leiden University (2014).
37 Visual acuity is not the only aspect that factors into a prescription, and the number given is an average for that acuity.
38 C. D. Andriesse, *Huygens: the man behind the principle* (Cambridge University Press, 2005) at p. 12.
39 J. S. Held, 'Constantijn Huygens and Susanna van Baerle: a hitherto unknown portrait', *Art Bull.* **73** (4), 653–668 (1991), at p. 656.

### Discussion and conclusions

Christiaan Huygens made a semi-empirical table that dictated the way that he and his brother would design telescopes. This table is seemingly based on a single optimized telescope from which Huygens used the telescope parameters and applied them to a different range of objective foci. This table was created when he was in his mid 50s, likely in 1684 or 1685, and it was updated at a later point to allow for about 10% more magnification, although possibly just to simplify the mathematics. These tables were accompanied with verbally constructed equations that closely approximated the tables and could be used when the latter were not at hand. However, the telescopes made by Huygens exhibit parameters that are far from being theoretically optimal, mostly because Huygens overmagnified his images and used very short eyepieces.

Since we know that Huygens was trying to make the optimal telescope (e.g. clearest image and strongest magnifying) and that he did not shy away from using extremely long aerial telescopes, it is not unreasonable to assume that he would have pushed his telescopes beyond this self-imposed limit if he knew that it was possible, or at least discuss it in his works.

This introduces the possibility that Huygens' eyes were the limiting factor in the construction of his telescopes, leading to suboptimal designs because he could not discern any improvements beyond this point. If we assume this to be the sole reason for the discrepancy between the Huygens equations and the theoretically derived ones, then we can find an upper limit for his visual acuity, and that he would have scored 20/70 on his eye exam. On average this corresponds to a prescription of $-1.5$ diopters. However, it is important to point out that no direct way exists to convert visual acuity to a prescription and that this number is only given indicatively.

Since we know that several of Huygens' family members had myopia, including his father, it is not unreasonable to assume that Huygens could also have inherited this trait to some extent. However, he would have worn glasses if his condition was very strong, which implies that he was only somewhat affected by it. This seems in line with our estimate. Moreover, assuming that these telescopes were designed for Huygens' imperfect eyesight provides a possible explanation as to why his telescopes never gained a large circulation outside of his family.

Investigations of this type could be repeated for others who developed similar empirical relations, or for those who built telescopes with a set focus. Additionally, a deeper investigation into Keplerian[41] telescopes by Huygens should be performed to see how well they adhere to his relations. Optical models[42] of such telescopes could also shed more light on whether any additional constraints are present that cannot be seen from the equations alone.

### Disclosure statement

The author declares that he has no known interests that might appear to affect his ability to present data objectively, in the form of financial interests, personal relationships or otherwise.

---

40  HUG 1, fols 108r–111v, inc. Si numerus pedum foci distantiae. (n.d.). Brill: https://doi.org/10.1163/2468-0303-cohu_01-092.
41  Telescopes that consist of two positive lenses.
42  Lozi *et al.*, *op. cit.* (note 33).

Data availability

The data and code used for this article are available in the project github (https://github.com/AlexPietrow/Huygens).

Acknowledgements

I thank Anders Nyholm, Gilles Otten, Huib Zuidervaart, Flavio Calvo, Göran Scharmer, Jan-Willem Pel and Kees Moddemeijer for the stimulating discussions on this topic, as well as Dominique Petit for her help with the translation of the Latin and French texts. Furthermore, I thank Malcolm Druett and Carsten Denker for their suggestions on improving the readability of the text, and the anonymous referees for their valuable suggestions during the peer review process.

I acknowledge NASA's Astrophysics Data System Bibliographic Services, Linda Hall Library, Bibliothèque nationale de France, Internet Archive and Google Books for making scanned versions of older literature freely available on the Internet.

I acknowledge the community effort devoted to the development of the following open-source packages that were used in this work: numpy (numpy.org), matplotlib (matplotlib.org), astropy (astropy.org).

I was supported by a grant for research infrastructures of national importance from the Swedish Research Council (registration number 2017-00625), the European Commission's Horizon 2020 Program under grant agreements 824064 (ESCAPE—European Science Cluster of Astronomy & Particle Physics ESFRI Research Infrastructures) and 824135 (SOLARNET—Integrating High Resolution Solar Physics).